\documentstyle[aps,prl]{revtex} 
\input epsf.sty 

\begin{document}
\renewcommand{\theequation}{\arabic{equation}} 

\twocolumn[ 
\hsize\textwidth\columnwidth\hsize\csname@twocolumnfalse\endcsname 
\draft 

\title{Higgs mechanism and superconductivity in U(1) lattice gauge
theory \\ with dual gauge fields 
}

\author{Tomoyoshi Ono,$^1$ Yuki Moribe,$^1$
Shunsuke Takashima,$^1$
Ikuo Ichinose,$^1$  Tetsuo Matsui,$^2$ and  
Kazuhiko Sakakibara$^3$ } 
\address{$^1$Department of Applied Physics,
Nagoya Institute of Technology, Nagoya, 466-8555 Japan 
}
\address{$^2$Department of Physics, Kinki University, 
Higashi-Osaka, 577-8502 Japan 
}
\address{$^3$Department of Physics, Nara National College of Technology, 
Yamatokohriyama, 639-1080 Japan
} 

\date{\today}

\maketitle 

\begin{abstract}   
We introduce a U(1) lattice gauge theory with dual gauge
fields and study its phase structure.
This system is motivated by
unconventional superconductors like extended $s$-wave
and $d$-wave superconductors in the strongly-correlated electron systems. 
In this theory, the ``Cooper-pair" field is put on links 
of a cubic lattice due to strong on-site repulsion between electrons
in contrast to the ordinary $s$-wave Cooper-pair field on sites. 
This Cooper-pair field behaves as  a 
gauge field dual to the electromagnetic U(1) gauge field.
By Monte Carlo simulations we study this lattice gauge model
and find a first-order phase transition from 
the normal state to the Higgs (superconducting) state.
Each gauge field works as a Higgs field for the other gauge field.
This mechanism requires no scalar fields in contrast to the ordinary
Higgs mechanism.
\end{abstract} 
]

\setcounter{footnote}{0} 

{\it Introduction. $-$ } 
The Ginzburg-Landau (GL) theory has proved itself a powerful
tool to  describe the phase transitions of conventional
$s$-wave superconductors. In field-theory terminology,
the GL theory takes a form of Abelian Higgs model (AHM),
and its phase structure has been studied by 
field-theoretical techniques and 
Monte Carlo (MC) simulations of lattice gauge theory.
These studies are partly motivated by
the work of Halperin, Lubensky, and Ma\cite{HLM} 
which predicts a first-order phase transition.
At present, it is established that the phase structure of 
three dimensional (3D) AHM on the lattice strongly depends on a parameter 
controlling fluctuations of the amplitudes $|\varphi(x)|$ 
of the Higgs (Cooper-pair) field\cite{AHM}.
At the London limit  in which $|\varphi(x)|$ is fixed, 
there is only the confinement phase in the lattice model.
As the fluctuations of $|\varphi(x)|$ are increased, 
a second-order phase transition to the Higgs phase appears, 
and for further fluctuations, the transition becomes of 
first-order.

Some strongly-correlated electron systems
exhibit unconventional superconductivity (UCSC)
\cite{ucsc} at low temperatures 
($T$). The first $d$-wave superconductor CeCu$_2$Si$_2$ was 
discovered in 1979\cite{Ce}.
In 1986, the cuprate high-$T_c$ superconductors were 
discovered\cite{highTc}, 
and later, it was found that they are $d$-wave superconductors.
Thus, it is interesting to set up and study the  GL theory of 
the UCSC. 
In the framework of weak-coupling theory,
such studies have appeared\cite{dwave}. However, 
the strong-coupling region remains to be 
studied\cite{twoHiggs}. 
In this Letter, we shall introduce a GL theory for 
the UCSC on a lattice, 
and study its phase structure by means of MC simulations.
We shall see that the order parameter, a bilocal field, 
is regarded as a gauge field, and
the knowledge and method of gauge theory are useful
to study this GL lattice gauge theory.
We find that this new type of gauge theory has a very interesting
phase structure.

{\it Lattice gauge model for UCSC. $-$  }
Let us first consider a UCSC on a 3D spatial lattice. 
We put a ``Cooper-pair field" $V_{xj}$ on the
link ($x, x+j$) of the lattice because of the strong on-site repulsion, 
where $x$ is the site index and
$j (=1,2,3)$ is the direction index 
(it also denotes the unit vector of the $j$-th direction).
$V_{xj}$ is related to electrons as
\begin{equation}
V_{xj}\propto\langle  C^\dagger_{x\uparrow}C^\dagger_{x+j,
\downarrow}
  -C^\dagger_{x\downarrow}C^\dagger_{x+j,\uparrow} \rangle,
\label{V}
\end{equation}
where $C_{x\sigma}$ is the electron operator at $x$ with spin 
$\sigma=\uparrow, \downarrow$.
In the rest of paper, we focus on the London limit of the 
$V_{xj}$ and put $|V_{xj}|=1$,
i.e., $V_{xj}=\exp(i\theta^v_{xj})$.
At present it is believed that in the underdoped region of the high-$T_c$ 
superconductors the SC phase transition is a sort of the Bose-Einstein 
condensation of the Cooper-pair field $V_{xj}$, i.e., $V_{xj}$ has a finite
amplitude and its phase fluctuation induces the SC phase transition.

There is another link field $U_{xj}$,
a compact U(1) gauge field, describing
the electromagnetic vector potential $\theta^u_{xj}$,
$
U_{xj}=e^{i\theta^u_{xj}}.
$
The original electron system is invariant
under a local gauge transformation $C^\dagger_{x\sigma} 
\rightarrow e^{i\varphi_x}C^\dagger_{x\sigma}$.
Under this transformation,
$V_{xj}$ and $U_{xj}$ transform as 
\begin{equation}
V_{xj}\rightarrow e^{i\varphi_{x+j}}V_{xj}e^{i\varphi_{x}},
\ \ U_{xj}\rightarrow e^{i\varphi_{x+j}}U_{xj}e^{-i\varphi_{x}}.
\label{gaugetr}
\end{equation}

By integrating over $C_{x\sigma}$ in path-integral method,
one obtains the action $A_{\rm GL}$ of the GL theory, 
which is expressed in terms of 
two U(1) gauge fields $V_{xj}$ and $U_{xj}$.
Because $A_{\rm GL}$ must respect the local gauge invariance 
under Eq.(\ref{gaugetr}), it is straightforward to ``derive" 
$A_{\rm GL}$  in the local expansion as 
\begin{eqnarray}
A_{\rm GL}&=&\frac{1}{2}\sum_{\rm pl}\Big[c_u U^4 +c_v V^4
+c_m (UVUV+VUVU)
\nonumber\\
&&+d_m (UUVV+{\rm 3\ permutations})\Big]+{\rm c.c.},
\label{AGL}
\end{eqnarray}
where $c_u$ etc. are effective parameters and some of them are
increasing functions of $1/T$. 
Each term in $A_{\rm GL}$ is 

\begin{figure} 
  \begin{picture}(0,105) 
    \put(30,0){\epsfxsize 170pt  
    \epsfbox{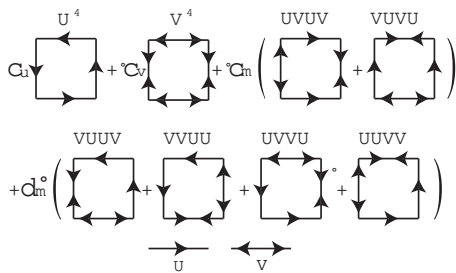}} 
  \end{picture}
  \caption{Action $A_{\rm GL}$ of the GL theory (\ref{AGL}). 
} 
\label{fig1}
\end{figure}

\noindent
depicted  in 
Fig.\ref{fig1}.
For example,  the $U^4$ term stands for the usual plaquette term 
$U^\dagger_{xi}U^\dagger_{x+i,j}U_{x+j,i}U_{xj}$ 
of lattice gauge theory.
The partition function $Z$ is given by
\begin{equation}
Z=\int [dU][dV]\exp(A_{\rm GL}),
\label{Z}
\end{equation}
where $[dU]\equiv \prod_{xj}d\theta^u_{xj}/2\pi$, etc.
We consider a 3D cubic
lattice of the size $L^3$ 
with the periodic boundary condition. 
(The results given below are for $L=16$ and $24$.)

For vanishing $c_m$ and $d_m$, the two gauge fields $U_{xj}$ and 
$V_{xj}$ decouple with each other and the system reduces to two 3D pure 
U(1) gauge systems.
As no phase transition takes place in the 3D U(1) 
pure gauge theory\cite{polyakov}, the present system for $c_m=d_m=0$ 
has only a single (confinement) phase.
Then we assign various nonvanishing values to $c_m$ and/or $d_m$, 
and determine the phase structure in the $c_u-c_v$ plane
for fixed $c_m$ and $d_m$.
We note that the system has a symmetry 
$Z(c_u,c_v,c_m,d_m)=Z(c_v,c_u,c_m,d_m)$, corresponding to
the interchange $U_{xj}\leftrightarrow V_{xj}$.
Below we present the results for two typical cases:
(i) $d_m = 0$ and $c_m > 0$ and (ii) $c_m = 0$ and $d_m < 0$.


{\it (i) Extended s-wave SC case ($d_m = 0$ and $c_m > 0$). $-$  }
To study the phase structure
we measure the internal energy 
$E$ and the specific heat $C$ (fluctuation of $E$),
\begin{equation}
E\equiv -\langle A_{\rm GL} \rangle/L^3 , \;\; 
C\equiv \langle (A_{\rm GL}-\langle A_{\rm GL} \rangle)^2\rangle/L^3.
\end{equation}
We considered $d_m=0, c_m=0.2, 0.4, 0.6, 0.8, 1.0,$ 
and $1.2$.

\begin{figure} 
  \begin{picture}(0,165) 
    \put(-10,95){\epsfxsize 115pt  
    \epsfbox{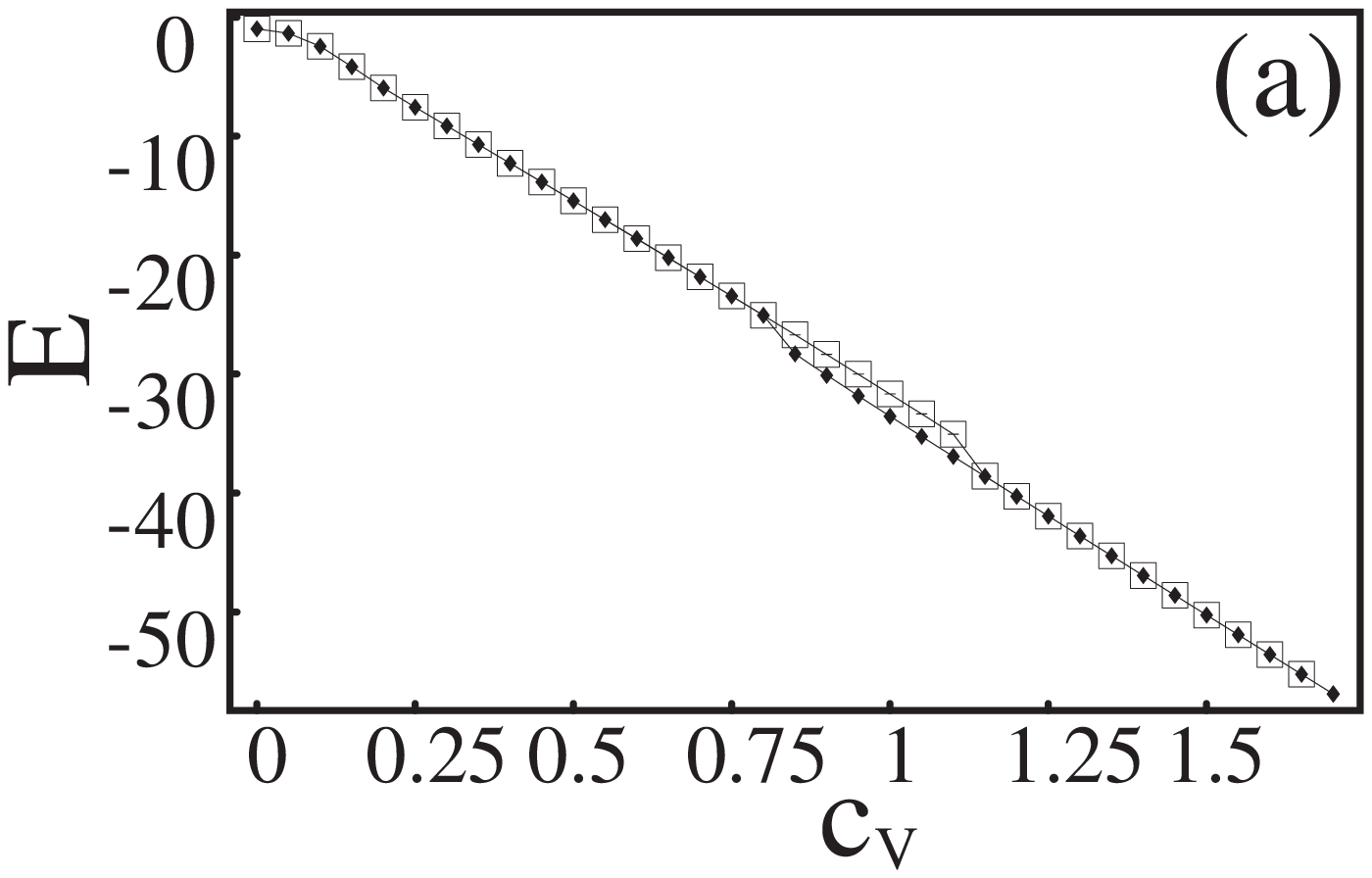}} 
    \put(110,95){\epsfxsize 115pt  
    \epsfbox{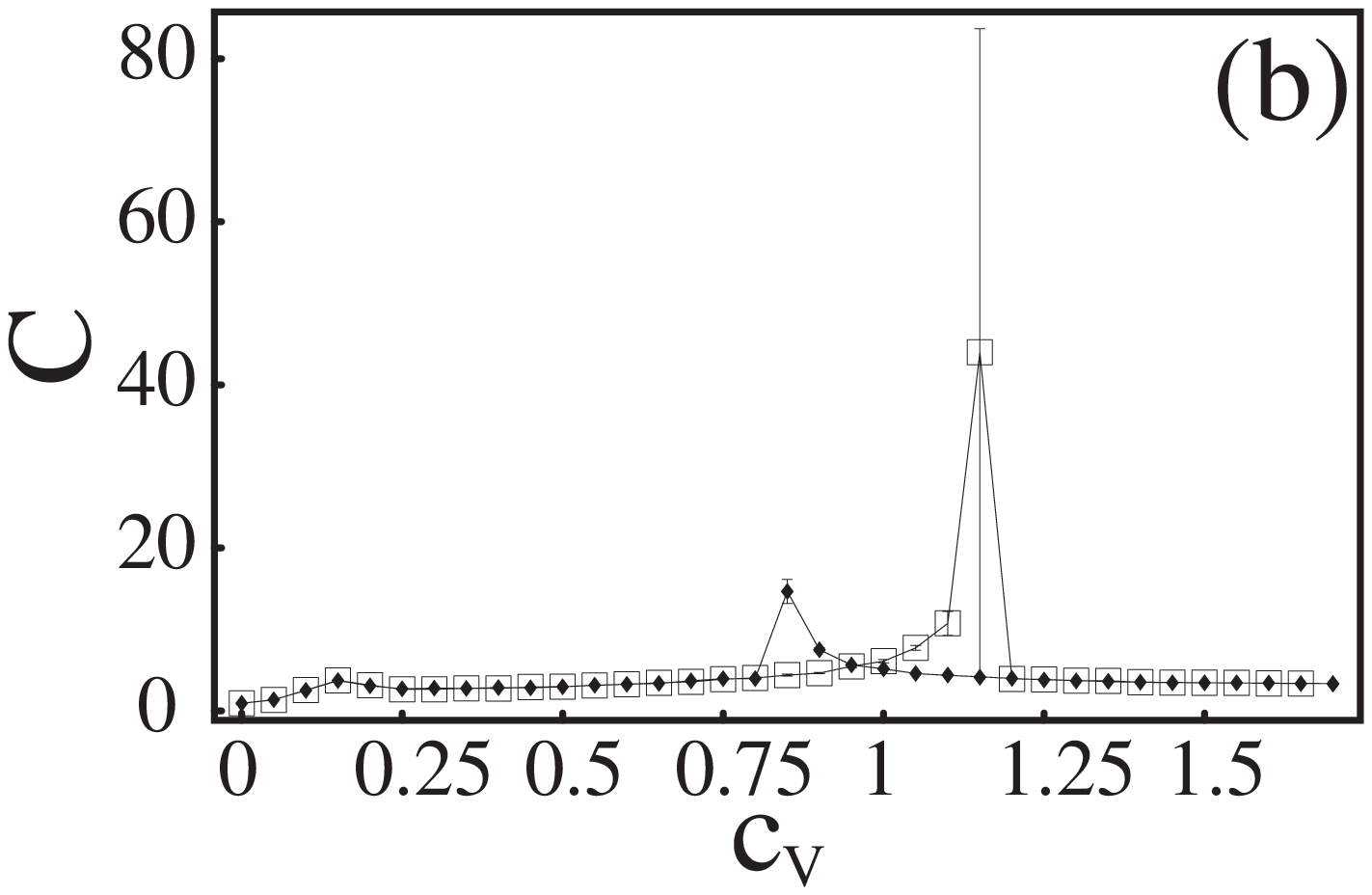}} 
    \put(10,1){\epsfxsize 95pt  
    \epsfbox{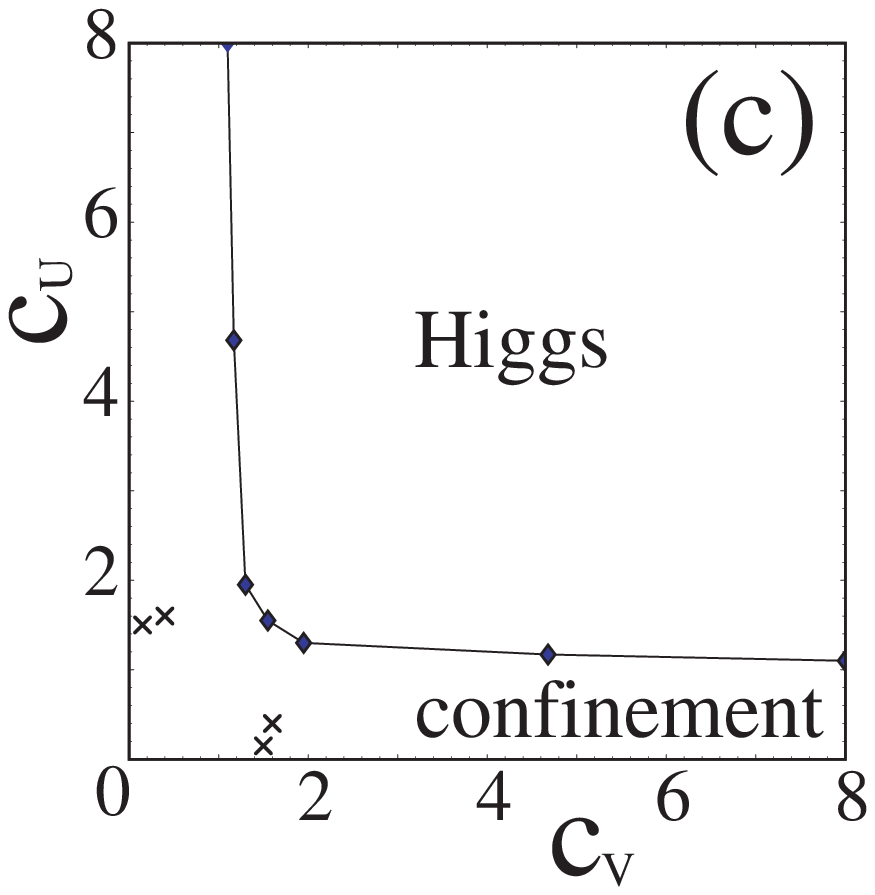}} 
    \put(110,5){\epsfxsize 120pt  
    \epsfbox{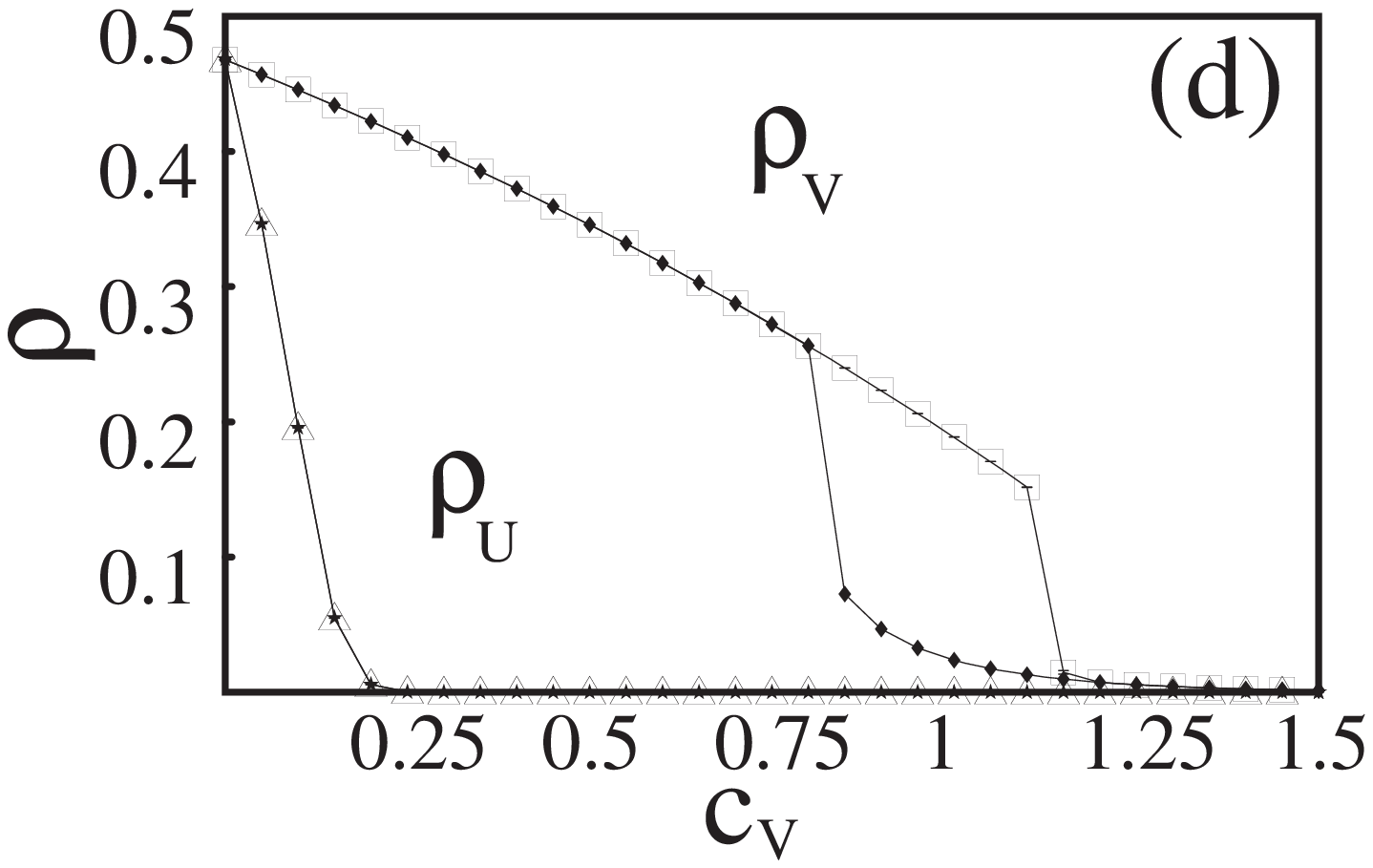}} 
  \end{picture}
\caption{Results for $d_m=0,c_m=0.6$.
(a) Internal energy $E$ for $c_v/c_u=0.1$
shows a hysteresis around $c_v \sim 1.0$,
a signal of first-order transition. 
(b) Specific heat  $C$ for $c_v/c_u=0.1$
shows a small cusp at $c_v\sim 0.17$,
 which is interpreted  as a crossover(see the text).
(c) Phase structure determined
by $E$ and $C$ for $L=16$.
The first-order transition line  
separates the confinement phase and the Higgs phase.
The cross symbols denote crossover.
(d) Instanton densities $\rho_U$ and $\rho_V$ 
for $c_v/c_u=0.1$. 
$\rho_V$ exhibits a discontinuity at $c_v\sim 1.0$,
while $\rho_U$ decreases rapidly at $c_v\sim 0.17$.
}
\label{fig2}
\end{figure}

In Fig.\ref{fig2}(a),(b) we show $E$ and $C$ for 
$d_m=0, c_m=0.6$ and $c_v/c_u=0.1$.
At $c_v \sim 1$,
$E$ shows a hysteresis, which implies a first-order phase transition.
The $c_m$ term works as a ``Higgs coupling" of the ``Higgs"
field $V_{xj}(U_{xj})$ to the gauge field 
$U_{xj}(V_{xj})$ to stabilize their fluctuations 
and induce such a transition\cite{CSS}. 
Thus the transition is expected from the confinement phase 
where $U_{xj}$ and $V_{xj}$ fluctuate violently to
the Higgs (superconducting) phase where their fluctuations are small. 

The data for $c_m \ge 0.6$ show signals of first-order 
phase transitions, while the data of $c_m \le 0.4$ 
show no signals of phase transitions.
In Fig.\ref{fig2}(c) we show the phase diagram in the $c_u-c_v$ plane
for $c_m=0.6$.
Similar phase diagram is obtained for $c_m=0.8\sim1.2$.

To confirm the above interpretation of each phase, 
we measured instanton densities.
We consider two kinds of
\noindent
instantons, i.e., $U$-instantons and 
$V$-instantons, and denote their average densities  per cube
as $\rho_U$ and $\rho_V$, respectively.
For the $U$-instantons we employ the definition 
given in Ref.\cite{inst},
which measures magnetic fluxes emanating from each smallest cube.
Similar gauge-invariant definition is possible for the
$V$-instantons\cite{Vinst}.
In Fig.\ref{fig2}(d) we plot $\rho_U$ and $\rho_V$.
$\rho_V$ shows a discontinuity 
at the first-order transition point $c_v\sim 1.0$ just like $E$, and 
decreases very rapidly as $c_v$ increases.
On the other hand, $\rho_U$ decreases very rapidly and almost vanishes for 
$c_v >0.2$.
This result indicates that the small 
cusp in $C$ of Fig.\ref{fig2}(b) at $c_v\sim 0.17$ 
reflects a crossover from the dilute to dense instanton ``phases"
of the gauge field $U_{xj}$\cite{polyakov,inst,TIM}.
Then we conclude that the system changes from
the confinement phase to the Higgs phase as $c_v \ (c_u)$ increases.

\begin{figure} 
  \begin{picture}(0,80) 
    \put(0,0){\epsfxsize 110pt  
    \epsfbox{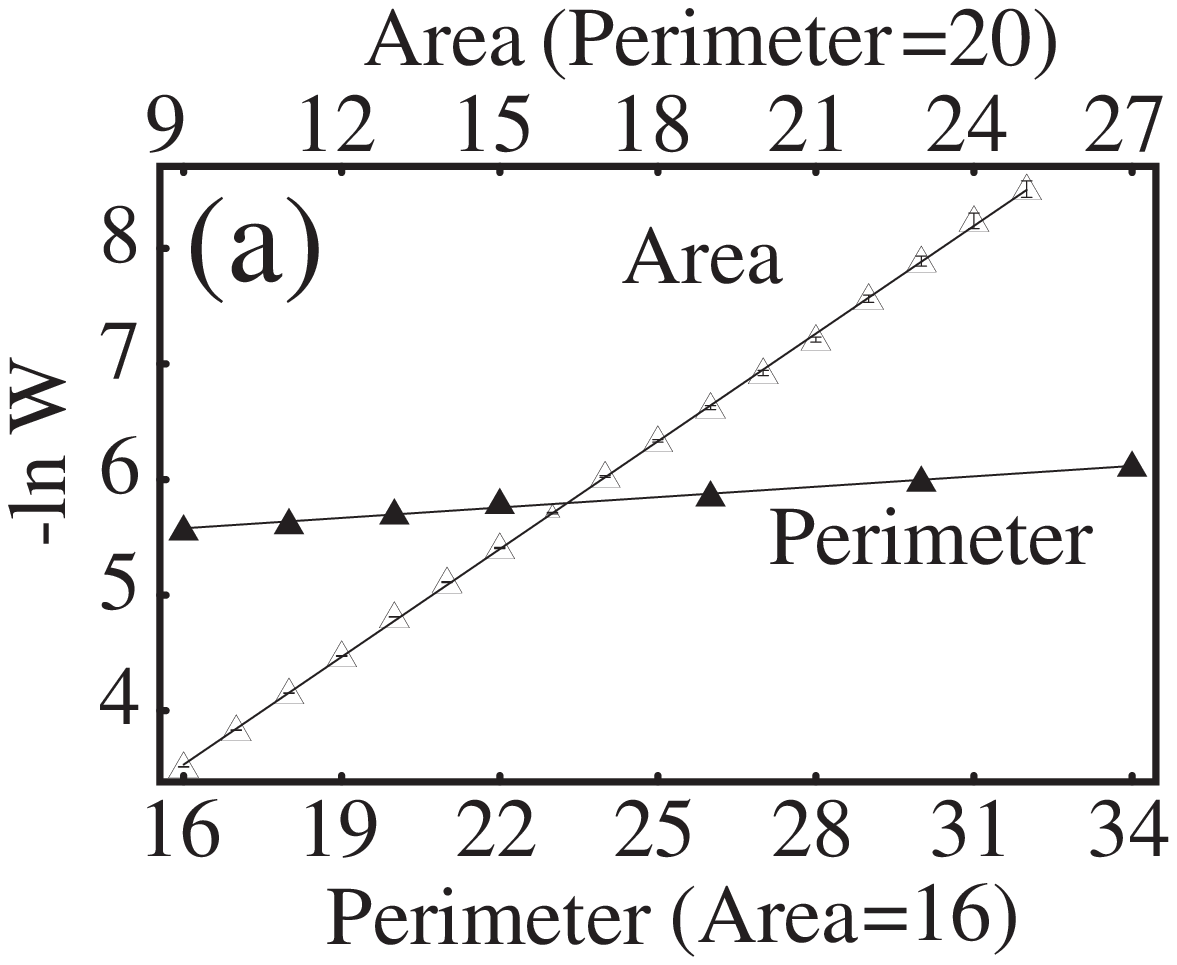}} 
    \put(120,0){\epsfxsize 115pt  
    \epsfbox{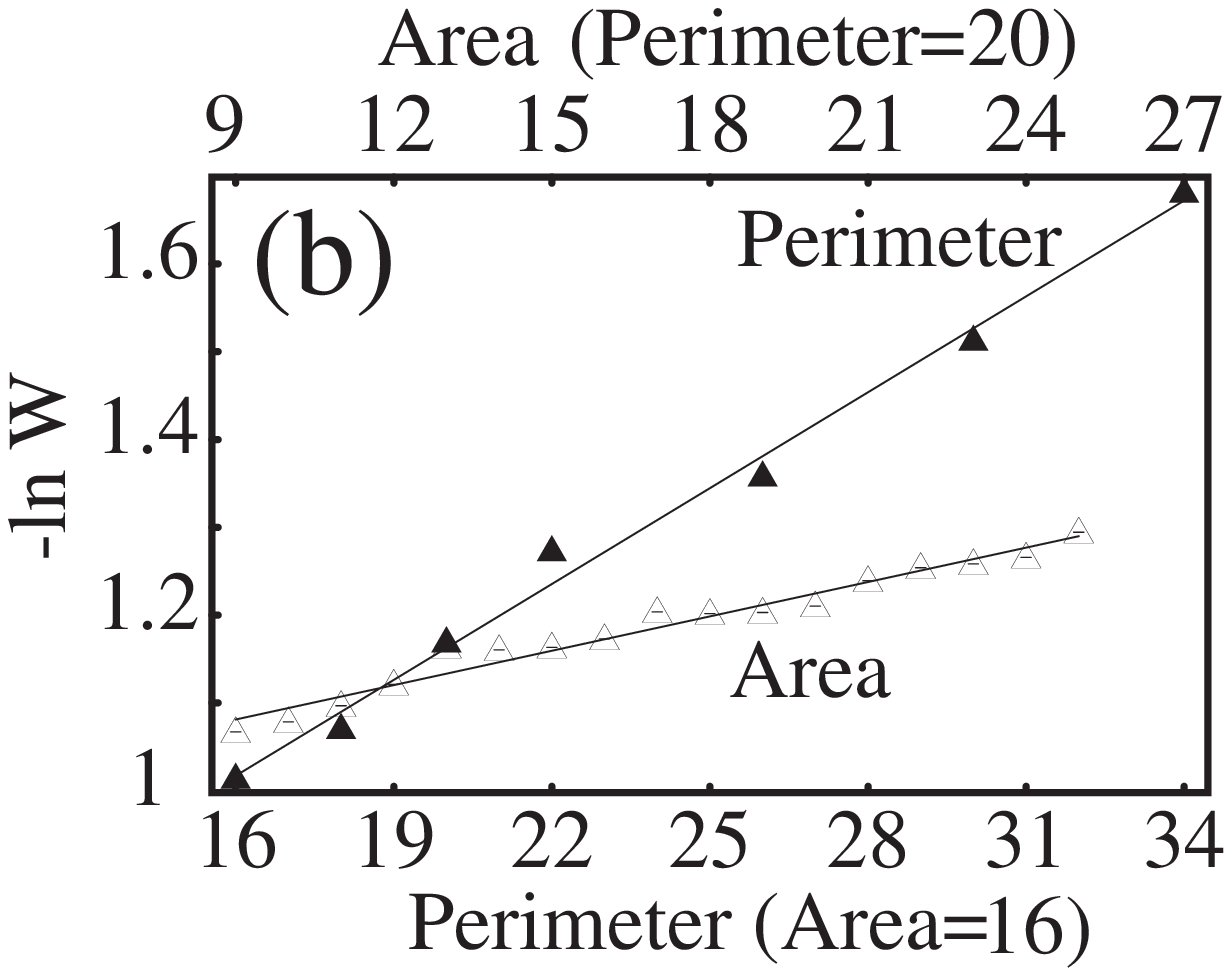}} 
  \end{picture}
\vspace{0.3cm}
\caption{Wilson loops for $d_m=0, c_m=0.6$ with
$c_u=c_v$, which means $W_U(\Gamma)=W_V(\Gamma)$.
(a) The case $c_u=c_v=1.4$ fits  
the area law.
(b) The case $c_u=c_v=2.0$ fits  
the perimeter law rather than the area law.
Fig.2(c) indicates that the transition between the two laws
occurs  at $c_u=c_v \sim 1.6$.}
\label{fig3}
\end{figure}

In order to support the above conclusions, 
we also calculated expectation values 
of the $U$ and $V$-Wilson loops,
\begin{equation}
W_U(\Gamma)=\langle \prod_\Gamma U \rangle, \;\; 
W_V(\Gamma)=\langle \prod_\Gamma V \rangle,
\label{Wloop}
\end{equation}
where $\Gamma$ is a closed loop on the lattice, and the products of 
$U_{xj}$ and $V_{xj}$ in Eq.(\ref{Wloop}) are formed to be gauge-invariant.
From the above results of instanton densities,
we expect that $W_{U(V)}(\Gamma)$ obey the are law in
the instanton-plasma 
phase for small $c_{u(v)}$
and the perimeter law in the instanton-dipole phase for large 
$c_{u(v)}$.
In calculating $W_{U(V)}(\Gamma)$, we consider various shapes of 
$\Gamma$.
For example, we take $\Gamma$'s having a fixed area
and various perimeters, and vice versa.
In Fig.\ref{fig3}
we show the results for $c_m=0.6$.
In the case $c_u=c_v=1.4$, $W_{U}(\Gamma)$ fits the
 area law,
while the case $c_u=c_v=2.0$ fits the perimeter

\noindent
 law.
These results support the previous conclusion obtained from 
instanton densities.

Because all the coefficients in $A_{\rm GL}$ are positive in the present case,
we expect that the observed SC state is an {\em extended} $s$-wave 
superconductor.
We calculated $\langle UUVV \rangle$ in order to verify this expectation
for $c_m\ge 0.6$ and $d_m=0$ cases and found that $\langle UUVV \rangle$ has
vanishing value in the normal state whereas it has a finite value in the Higgs 
phase showing hysteresis loop.
Furthermore its value takes a negative as well as positive value 
depending on samples.
Turning on a small but finite positive $d_m$ term, $\langle UUVV \rangle$ 
becomes positive. 
This result indicates that positive $d_m$ term is necessary to produce
a genuine {\em extended} $s$-wave superconductor.


{\it (ii) d-wave SC case ($c_m = 0$ and $d_m < 0$). $-$  }
We expect and verified that the $d_m$-term with $d_m < 0$ 
enhances $d$-wave condensation of $V_{xj}$;
$ \Delta_{ij} \equiv
\langle U_{xi}U_{x+i,j}V^\dagger_{x+j,i}V_{xj} \rangle 
< 0$. For large $c_u$, 
the fluctuations of $U_{xj}$ are suppressed 
as $U_{xj}\sim 1$ [up to Eq.(\ref{gaugetr})],
hence the $d$-wave configuration 
$\langle V^\dagger_{xi}V_{x+i,j}\rangle < 0$ 
$(i\neq j)$ is preferred, although there are no configurations 
with all negative $\Delta_{ij}$.
We considered the cases $d_m=-0.4,\;-0.6, \; -0.8, \; -1.0$, and
measured $E,\; C$,  $\rho_U,$ and $\rho_V$ as before.
No signals of phase transitions are found for $d_m=-0.4$ and $-0.6$,
whereas signals of phase transitions to the 
$d$-wave superconducting phase are obtained for $d_m <  -0.6$.

\begin{figure} 
  \begin{picture}(0,170) 
    \put(-10,93){\epsfxsize 115pt  
    \epsfbox{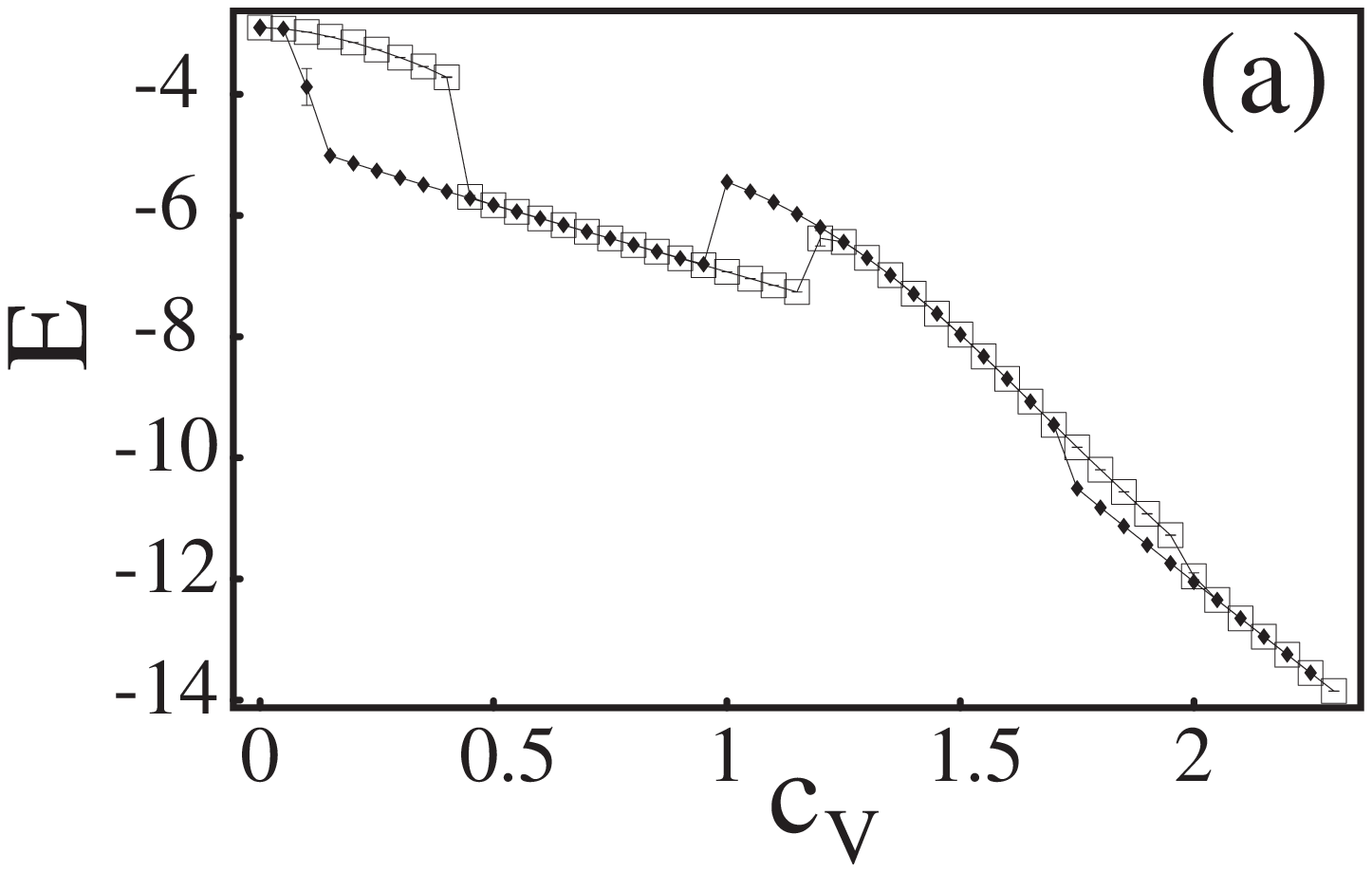}} 
    \put(105,93){\epsfxsize 115pt  
    \epsfbox{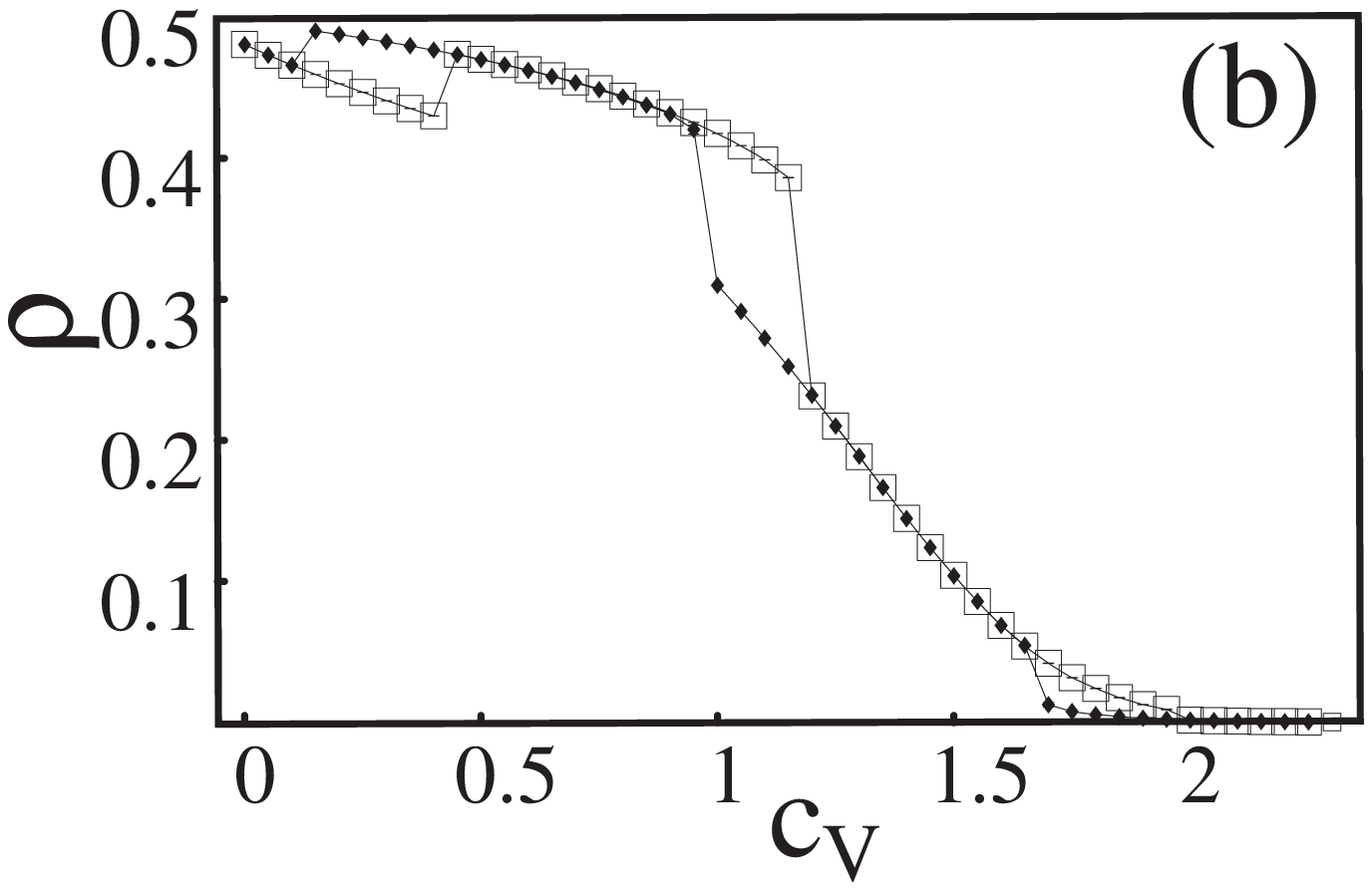}} 
    \put(1,-1){\epsfxsize 93pt  
    \epsfbox{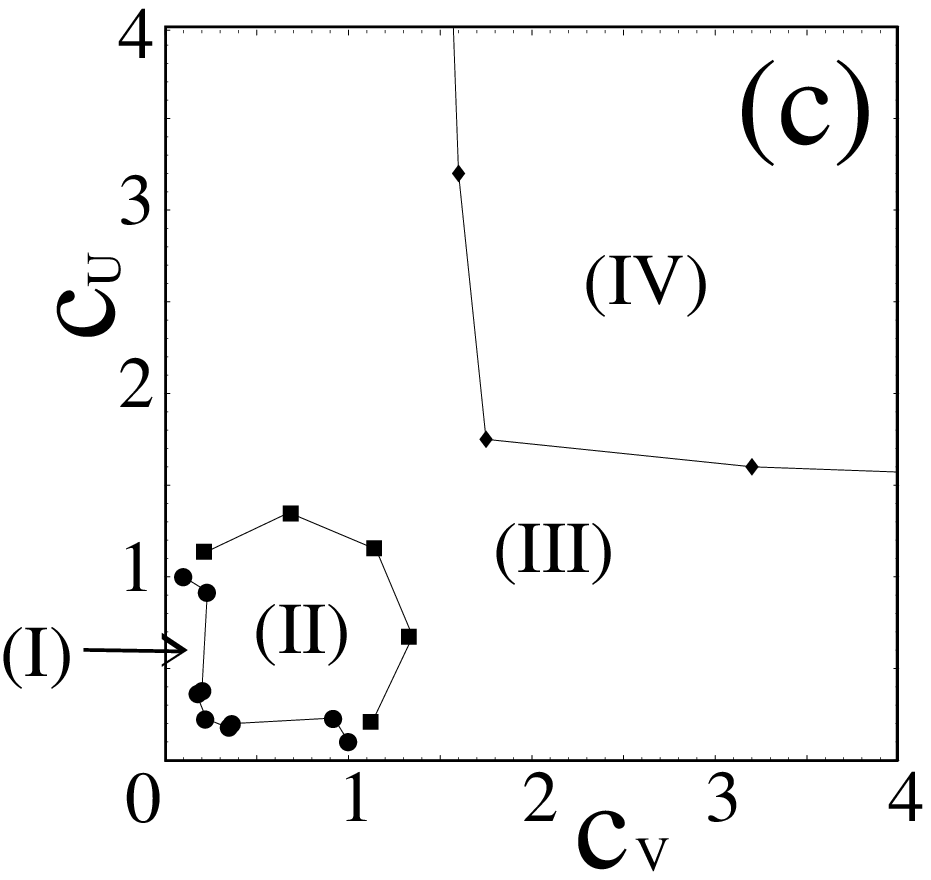}} 
    \put(110,0){\epsfxsize 110pt  
    \epsfbox{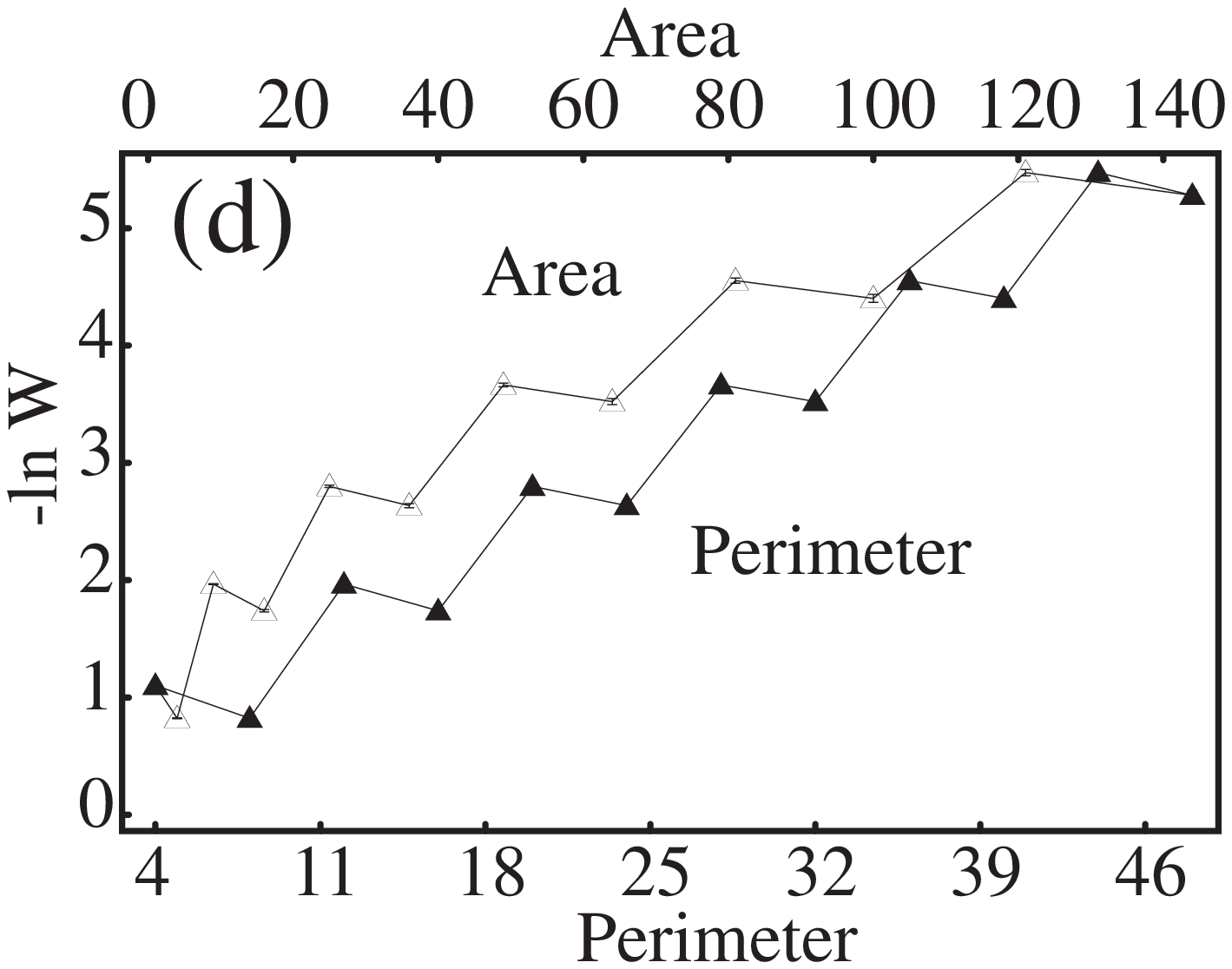}} 
  \end{picture}
  \vspace{0.5cm}
\caption{Results for $c_m=0, d_m = -0.8$. 
(a) $E$ and  (b) $\rho_U = \rho_V$ for $c_u/c_v=1.0$. 
They show three hysteresis loops. 
(c) Phase structure in the $c_u-c_v$ plane.
(d) Wilson loop of $M\times M$ square for $c_u= 
c_v=0.9$ shows two smooth branches for even $M$ and odd $M$, i.e.,
staggered perimeter law.
 }
\label{fig4}
\end{figure}

In Fig.\ref{fig4}(a),(b) we present $E$ and $\rho$ for $c_u/c_v=1.0$
and $d_m=-0.8$.
$E$ and $\rho$ show three first-order phase transitions along $c_v=c_u$.
In Fig.\ref{fig4}(c) we present the phase structure.
There are four phases (I)-(IV).
The phase (I) is the confinement phase. 
(II) is the ``staggered state"  which
is generated by the frustration of strong negative $d$-term.
It breaks the translational symmetry by the unit lattice spacing
as 
supported by the Wilson loop of Fig.\ref{fig4}(d)
which has two branches, one for even areas and one for odd areas. 
(III) is the disorder state connecting (II) and (IV).
(IV) is the Higgs phase corresponding to $d$-wave superconductor.
These interpretations are consistent  with the behavior of 
$\rho$ in Fig.\ref{fig4}(b).
The measurement of $\Delta_{\mu\nu}$ shows that the cubic symmetry 
$\Delta_{23} =\Delta_{12}=\Delta_{13}$ is maintained in (IV),
whereas it is reduced to the square

\noindent
symmetry, e.g., 
$\Delta_{23} < \Delta_{12}=\Delta_{13}$ in (II).

\begin{figure} 
  \begin{picture}(0,75) 
    \put(-5,0){\epsfxsize 115pt  
    \epsfbox{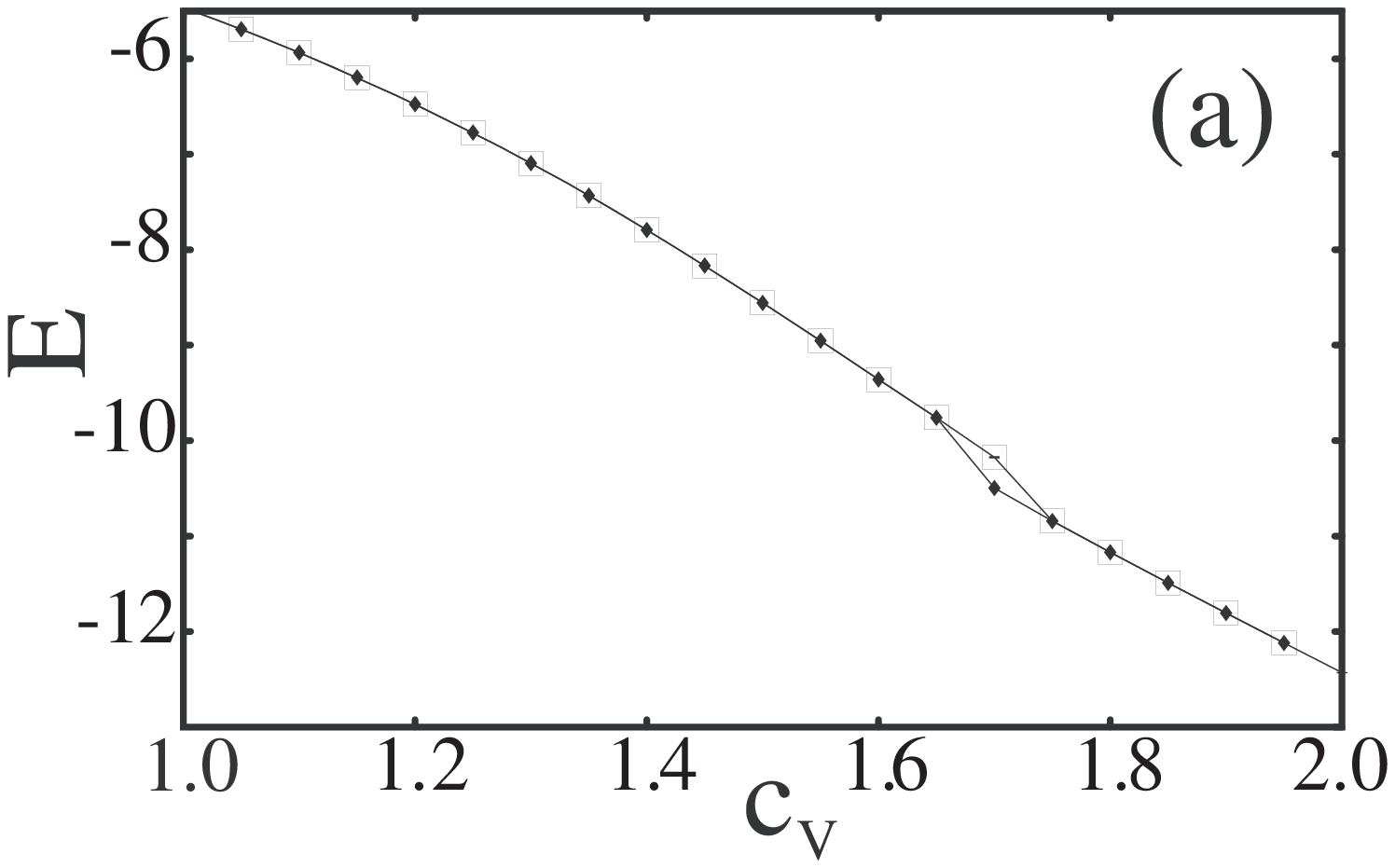}} 
    \put(112,0){\epsfxsize 120pt  
    \epsfbox{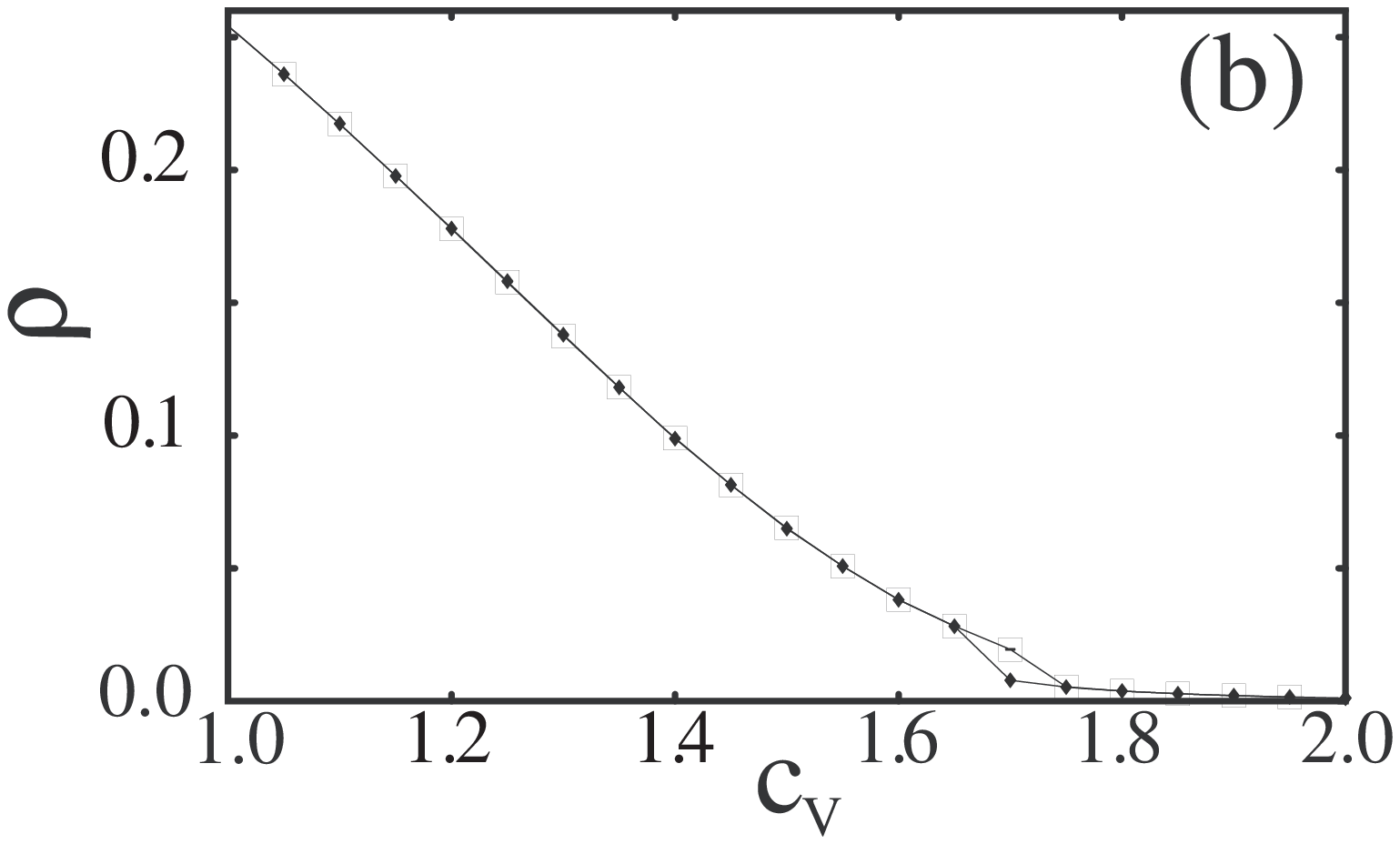}} 
  \end{picture}
\vspace{0.3cm}
\caption{(a)Internal energy for an anisotropic system. Parameters are
$c_u/c_v=1, \; c_m=0$, intraplane $d_m=-1.0$ and interplane $d_m=-0.8$.
There is a first-order phase transition near $c_u=c_v=1.7$.
(b)Instanton density also shows hysteresis loop near $c_u=c_v=1.7$.
}
\label{fig5}
\end{figure}

We also studied an anisotropic system in which the intraplane
$d_m$ term is larger than the interplane $d_m$ term because many
of the real materials have a layered structure.
In Fig.\ref{fig5} we show $E$ and the instanton density for 
$c_u/c_v=1.0, \; c_m=0$, intraplane $d_m=-1.0$ and interplane $d_m=-0.8$.
There is a first-order phase transition near $c_u=c_v=1.7$,
whereas the other transitions existing in the isotropic case in Fig.\ref{fig4}
disappeared. Thus the phase (I) and (II) disappear 
whereas the two phases (III) and (IV) survive.
This phase transition in the anisotropic case should correspond to 
the SC transition observed in the heavy-fermion materials, etc.

{\it Quantum phase transition. $-$ }
We have considered the gauge model $A_{\rm GL}$ (\ref{AGL}) defined on 
the cubic lattice  and studied its phase structure.
The phase transitions which we found in the previous discussion
correspond to {\em thermal phase transitions}, i.e., the coefficients
in $A_{\rm GL}$ are increasing functions of $1/T$ 
and  the SC phase appears as $T$ is lowered.
Recently, in the studies of the 
strongly-correlated electron systems  
like the high-$T_c$ cuprates and the heavy-fermion materials, 
significance of {\em quantum} 

\noindent
{\em phase transition} (QPT) has been 
recognized\cite{QPT}.
In particular, the QPT in ordinary $s$-wave charged SC was studied by using a  
XY model coupled with a U(1) gauge field\cite{fisher} and very recently 
this model was applied for the QPT in the 
high-$T_c$ cuprates in order to explain the anormalous behavior of 
the superfluid density near the quantum critical point (QCP) at $T=0$
\cite{franz}.

Quantum theory of the present GL theory can be constructed  
straightforwardly.
In the continuum imaginary-time formalism, 
$U_{xj}$ and $V_{xj}$ 
depend on  the imaginary time $\tau$.
Then we discretize the imaginary-time axis and define the quantum GL 
theory on the 4D hypercubic lattice.
The action  $A_{\rm GL}^{\rm q}$ of the quantum system
has a similar form of $A_{\rm GL}$
in Fig.\ref{fig1}, but it is defined on the 4D lattice.
It involves the time component $U_{x\tau}$
but does {\em not}
contain the terms including $V_{x\tau}$ 
as the Cooper-pair field lives
on links of the {\em spatial} lattice.

\begin{figure} 
  \begin{picture}(0,100) 
    \put(-5,5){\epsfxsize 115pt  
    \epsfbox{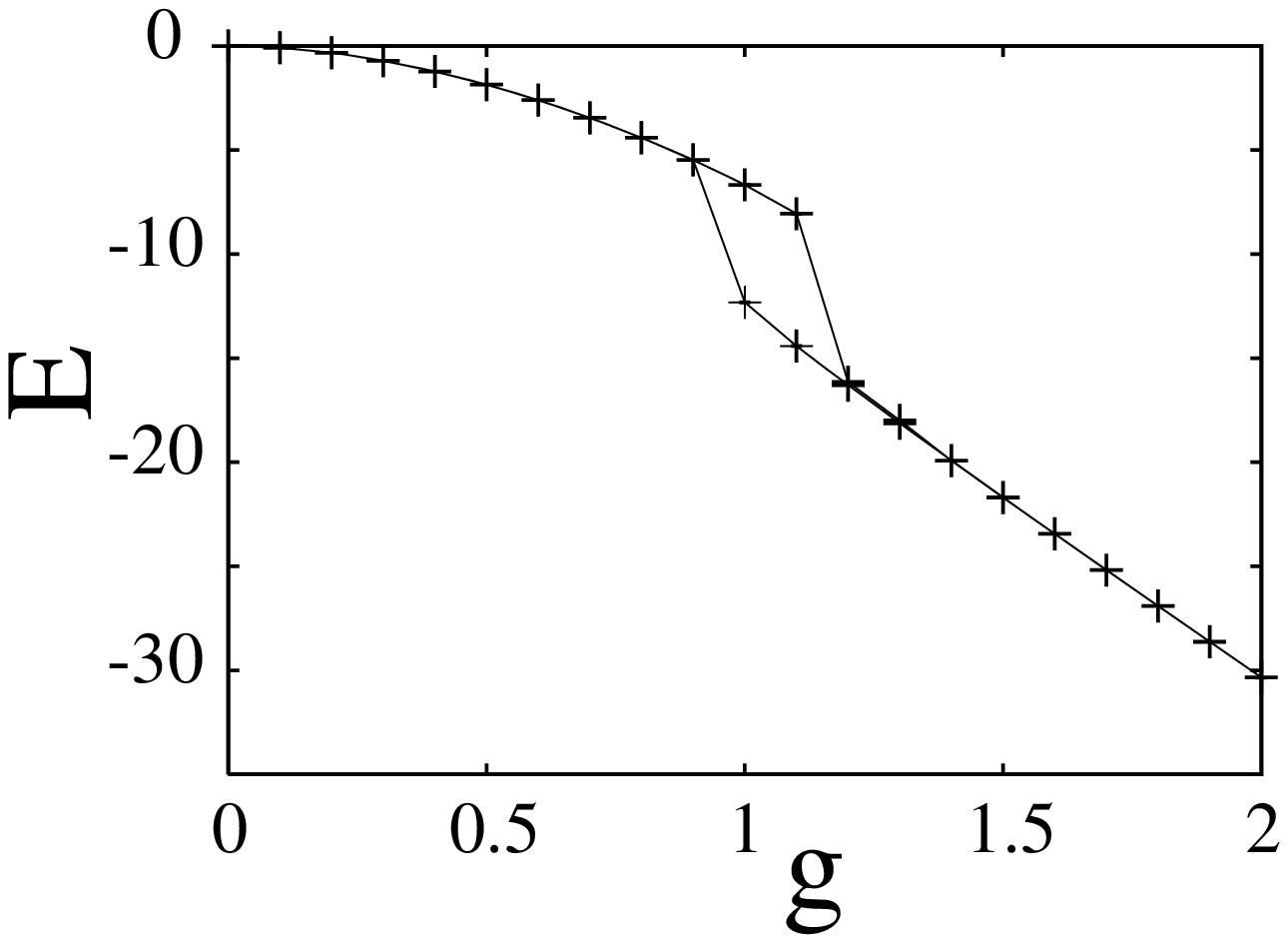}} 
\put(120,5){\epsfxsize 110pt  
    \epsfbox{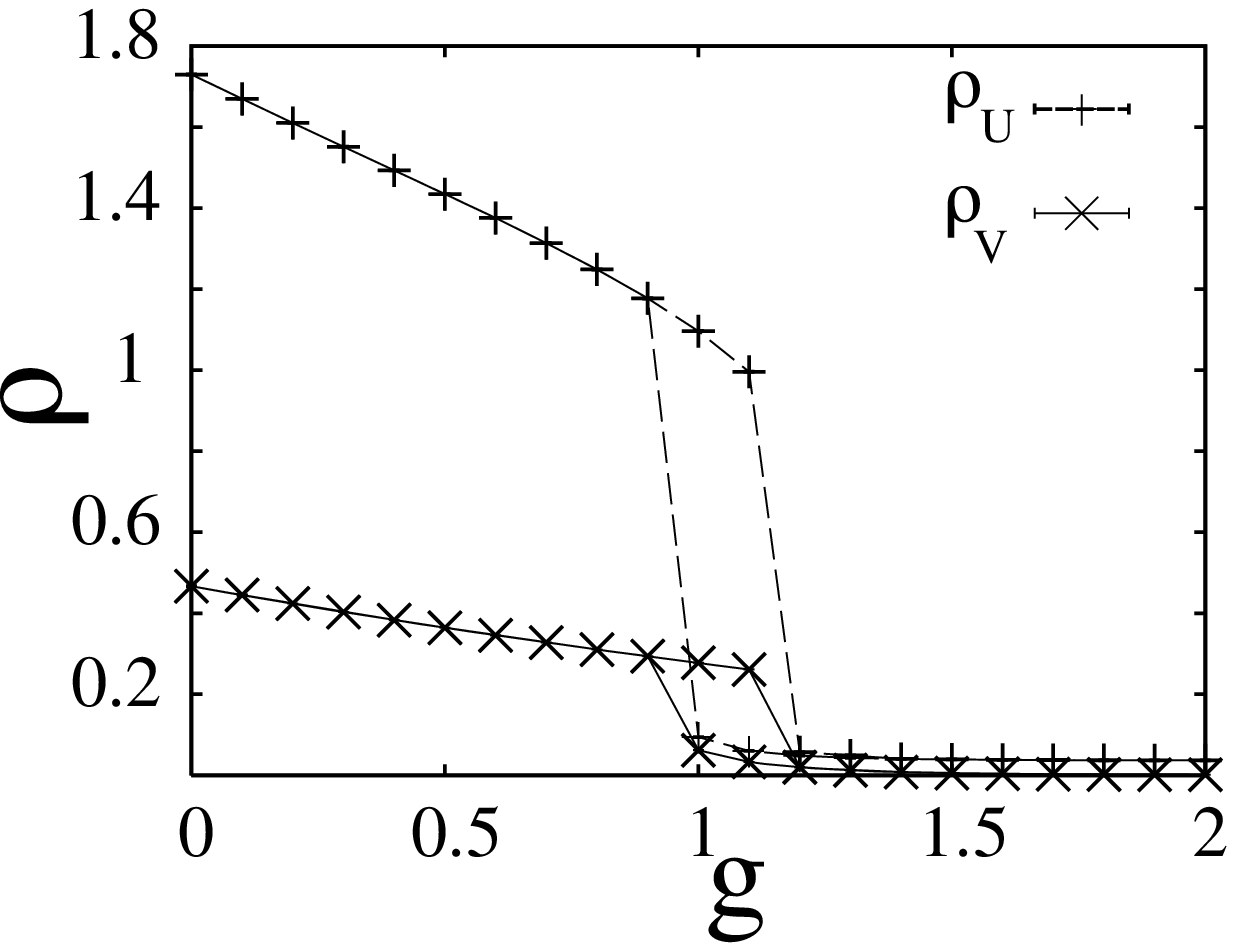}} 
  \end{picture}
  \caption{$E$ and instanton density of the quantum system
$A_{\rm GL}^{\rm q}$ with $(c_u,c_v,c_m,d_m)=g(1,1,1,0)$. 
They exhibit first-order phase transition.
} 
\label{fig6}
\end{figure}

We also studied 
this quantum system $A_{\rm GL}^{\rm q}$ by the MC simulations.
In the practical experiments, external conditions and 
properties of samples like external pressure, doping parameter, 
etc., change the effective parameters contained
in $A_{\rm GL}^{\rm q}$.
Here we consider typical two cases in which the coefficients in 
$A_{\rm GL}^{\rm q}$ are scaled as $(c_u,c_v,c_m,d_m)=g(1,1,1,0)$ and 
$g(1,1,1,-1)$ where $g$ is a positive parameter.
In Fig.\ref{fig6}, we show $E$ and the instanton density
for the former case.
The result shows a first-order phase transition
from the normal (or insulating) phase to the SC phase.
We found that the other case also has a similar phase structure.
This result should be compared with the 
observed SC phase transition 
at very low $T$'s in various heavy-fermion materials, etc.

{\it Conclusion. $-$  }
In this paper, we considered a GL theory for UCSC by introducing 
a ``Cooper-pair" field on spatial links.
We showed that the GL theory can be regarded a new type of lattice
gauge model which contains two kinds of gauge fields.
By means of MC simulations, we clarified its phase structure.
For the case of $c_m, d_m>0$, there are 
two phases for sufficiently large $c_m$, which correspond 
to the normal and SC (Higgs) phases.
They are separated by a first-order phase transition line.
On the other hand, for 
the case of $d_m \le -0.8$
there are four phases in the isotropic case.
Only the two of them

\noindent
 survive in the anisotropic 
case which corresponds to the layered structure.
The observed phase transitions should be compared with the experiments
of the high-$T_c$ cuprates and heavy-fermion materials.




\begin{references} 


\bibitem{HLM}B.I. Halperin, T.C. Lubensky, and S. Ma, 
Phys. Rev. \\ Lett. {\bf 32}, 292(1972). See also,
S. Coleman and E. Weinberg, Phys. Rev.{\bf D7}, 1888(1973).

\bibitem{AHM}
M.N. Chernodub, E.-M. Ilgenfritz, 
and A. Schiller, Phys. Lett. {\bf B547}, 269(2002); S. Wenzel, 
E. Bittner, W. Janke, A.M.J. Schakel, and A. Schiller, 
Phys. Rev. Lett.{\bf 95}, 051601(2005), and references cited therein.

\bibitem{ucsc}
V.P. Mineev and K.V Samokhin, 
``Intorduction to Unconventional Superconductivity"
(Gordon and Breach Scienece Publishers, 1999). 

\bibitem{Ce}F. Steglich, J. Aarts, C.D. Brell, W. Lieke, \\
D. Meschede, W. Franz, and H. Schafer, Phys. Rev. Lett.{\bf 43},
1892(1979).

\bibitem{highTc}
J.G. Bednorz and K.A. Mueller, Z. Phys.{\bf B64}, 189(1986).

\bibitem{dwave}
M. Sigrist and K. Ueda, Rev. Mod. Phys.{\bf 63}, 308(1993).

\bibitem{twoHiggs}Recently,
M.N. Chernodub, E.-M. Ilgenfritz, and A. Schiller, 
Phy. Rev. {\bf B73}, 100506(2006),
studied a related gauge theory, where
a doubly-charged scalar field on sites 
is introduced instead of the $V$-field in the present paper.

\bibitem{polyakov}A.M. Polyakov, Nucl. Phys.{\bf B120}, 429(1977).

\bibitem{CSS}I.Ichinose and T.Matsui,
Phys.Rev.{\bf B51}, 11860(1995).

\bibitem{inst}T.A. DeGrand and D. Toussaint,
Phys. Rev.{\bf D22}, 2478 \\
(1980);
R.J. Wensley and J.D. Stack, Phys. Rev. Lett.{\bf 63}, 1764(1989). 

\bibitem{Vinst}
To define the $V$-instantons we consider the replacement
$\varphi_x \rightarrow -\varphi_x$  
for the odd sites
in Eq.(\ref{gaugetr}).
Then $V_{x\mu}$ transforms just like $U^\dagger_{x\mu}$ before the
replacement, so it is straightforward to define the $V$-instantons.

\bibitem{TIM} S. Takashima, I. Ichinose, and T. Matsui, 
Phys. Rev.{\bf B72}, 075112(2005).

\bibitem{QPT}S.Sachdev,
{\it ``Quantum Phase Transitions"} (Cambrige, \\
1999).

\bibitem{fisher}M.P.A.Fisher and G.Grinstein,
Phys.Rev.Lett.{\bf 60}, \\
208(1988).

\bibitem{franz}M.Franz and A.P.Iyengar,
Phys.Rev.Lett.{\bf 96}, \\
047007(2006).  



\end{references}
\end{document}